\documentclass{ifacconf}

\usepackage{graphicx}      
\usepackage{natbib}        

\usepackage[dvips]{epsfig}    

\usepackage{xcolor}

\usepackage{amssymb}
\usepackage{amsmath}          
\usepackage{enumerate}

\begin{document}
\begin{frontmatter}

\title{Power-based control of output oscillations with online estimation of biased harmonics}

\thanks[]{\textcolor[rgb]{0.00,0.00,1.00}{Authors accepted manuscript}}

\author[1]{Michael Ruderman},
\author[2]{Denis Efimov}


\address[1]{University of Agder, 4879 Grimstad, Norway}
\address[2]{Inria, University Lille, CNRS, 9189-CRIStAL, Lille, France  \\
Emails: \tt\small michael.ruderman@uia.no,  denis.efimov@inria.fr}

\begin{abstract}
The recently introduced discrete power-based control
(\cite{ruderman2024}) reduces largely the communication
efforts in the control loop when compensating for the marginally
damped or even slowly diverging output oscillations. The control
commutates twice per oscillations period (at the amplitude peaks) and uses the measured
harmonic output only. The power-based control scheme
requires the knowledge of the instantaneous frequency, amplitude,
and bias parameters of the harmonic signal. This paper extends the
power-based control by the finite-time estimation of the biased
harmonics (\cite{ahmed2022}). Also an improved analytic
calculation of the impulse weighting factor is provided. The
power-based oscillations control with online estimation of the
harmonic parameters is evaluated experimentally on the fifth-order
actuator system with a free hanging load under gravity and
measurement noise.
\end{abstract}


\end{frontmatter}


\section{Introduction}
\label{sec:1}

The control, or often just a mitigation, of the harmonic oscillations and/or vibrations in the output state is an essential issue for various mechanical systems and application scenarios. Examples can be found in e.g. active suspensions (\cite{landau2005adaptive}), off-shore rotary drilling systems (\cite{wang2020adaptive}), actuator drives with elastic connections (\cite{ruderman2012observer}), two-rotor vibrational systems (\cite{fradkov2020control}), to mention here a few. The associated control problems are often exacerbated by the fact that sensor and actuator elements are not co-located, like in case of e.g. drill string systems (\cite{aarsnes2018dynamics}) or large flexible structures (\cite{gibson2011modeling}). The non-collocated oscillatory systems can suffer under a limited access to the internal dynamic states and a poor observability, that makes an observer design and full-state feedback control hardly achievable, as demonstrated in \cite{ruderman2024adaptive}. If a non-collocated sensor/actuator configuration is in place, a reduced communication rate in the closed control loop can also be desired, cf. \cite{ruderman2024}. Such control with the reduced sampling or event-triggered may be advantageous in terms of the hardware components and communication architecture, but at the same time requires a reliable online estimation of the oscillation process parameters on the sensor side.

The problem of online estimation of parameters from a
linear regression is classical and has many solutions depending on
excitation of the regressor and the properties of measurement
perturbations \cite{Ljung1987,Sastry1989,Astolfi2008}. A usual
standing assumption requires the persistence of excitation
(\cite{Morgan1977}) of the regressor for robust estimation in face
of disturbances, while in the noise-free setting the necessary and
sufficient condition is the interval excitation. The most popular
solutions include the least squares algorithm or the gradient
descent one, which guarantee an exponential convergence of the estimates to their
ideal values under a sufficiently excited
regressor (\cite{Narendra1987}), while for a weaker excitation
the asymptotic convergence rates can be recovered, see
\cite{Efimov_PE2018}.

In the case when some parameters appear in a nonlinear fashion in
the regression, there are methods oriented on estimation for
convex/concave functions of the parameters (\cite{fomin1981adaptive}) or
other special scenarios (\cite{Tyukin2007,Ortega2024}). Recent
approaches use some preliminary algebraic operations reducing the
problem to a linear regression with respect to other parameters,
see e.g. \cite{Ushirobira2023}. A notable example here is the
problem of estimation of a frequency in a measured biased harmonic
signal, which has many solutions
\cite{Pin2019,vediakova2020frequency,ahmed2022,ruderman2024adaptive}, to mention a
few.

If the estimation algorithm represents a part with online control
loop, then often stricter requirements on the rate of convergence
and robustness are imposed. It has been observed, that in
conventional approaches it is difficult to accelerate since their
performance characteristics are governed by the level of
excitation, \cite{Efimov2015}, which may be hard to change.
Recently, several alternative solutions have been proposed, which
however require more computational power
\cite{Chowdhary2012,Aranovskiy2017}, but they can provide
finite/fixed-time convergence rates (\cite{efimov2021}) for the
estimation errors in linear regressions (\cite{Rios2017,Wang2020}). An example of the practical control scenario necessitating a rapid and reliable frequency estimation is the problem of  compensating mechanical oscillations of a passive inertial load, that was recently addressed in \cite{ruderman2024adaptive,ruderman2024}. Based thereupon, the present work combines the power-based control (\cite{ruderman2024}) with the finite-time estimation of the biased harmonics (\cite{ahmed2022}). The control scheme requires the knowledge of the instantaneous frequency, amplitude, and bias parameters of the output harmonic signal to be compensated. 

The rest of the paper is organized as follows. In section \ref{sec:2}, we resume the discrete power-based control (\cite{ruderman2024}) and provides the impulse weighting factor by analytic calculation of the amplitude reduction through the feed-forward sub-dynamics of the plant. The estimation of biased harmonics (\cite{ahmed2022}) is summarized in section \ref{sec:3}, while its convergence prosperities and parameters tuning are addressed in section \ref{sec:4}. A detailed experimental case study of the oscillations compensation with online harmonics estimation is shown in section \ref{sec:5}. The conclusions are derived at the end in section \ref{sec:6}.

\section{Power-based oscillations control}
\label{sec:2}

\begin{figure}[!h]
\centering
\includegraphics[width=0.8\columnwidth]{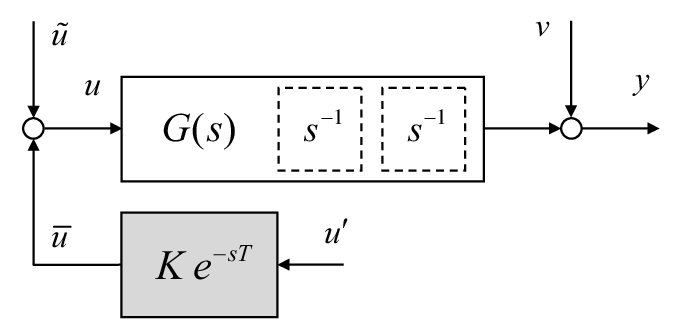}
\caption{Discrete power-based control $u^{\prime}$ subject to the gain
shaping and time-delay synchronization.}
\label{fig:controlstruct}
\end{figure}

The power-based oscillations control (\cite{ruderman2024}) is
applicable as a plug-in compensator of the oscillatory output quantity
$y(t)$, see Fig. \ref{fig:controlstruct}, while the overall plant
transfer function $G(s)$ assumes the double integrator in series
at the output channel. The corresponding dominant
conjugate-complex pole pair at $\lambda_{1,2} = \sigma \pm j
\omega$ is assumed to be either undamped i.e. $\sigma = 0$ or very low
damped i.e. $0 \leq \zeta \ll 1$; $\zeta = - \sigma
(\sigma^2 + \omega^2)^{-1/2}$ is the standard (normalized) damping
ratio. Note that the power-based control is also capable to
compensate for even slowly diverging output oscillations, i.e. for
$-1 \ll \zeta < 0$. This case will be demonstrated in the
experimental study where a proportional-integral (PI)
feedback controller $\tilde{u}$, cf. Fig. \ref{fig:controlstruct},
forces the otherwise stable $G(s)$ to have the closed-loop system
with a dominant pole pair with $\sigma > 0$.

It is assumed that the single measured output of the process has the
form
\begin{equation}
y(t)=Y_{0}+A\sin(\omega t+\phi)+v(t), \quad t\geq0,\label{eq:output}
\end{equation}
where $y(t),v(t)\in\mathbb{R}$ are the measured signal with the respective
bounded noise. $A>0$ and $\omega>0$ are the amplitude and the frequency
of oscillations, correspondingly, $Y_{0}\in\mathbb{R}$ is the bias
and $\phi\in[0,2\pi)$ is the phase. For the purpose of estimation,
we assume that all these parameters are quasi-constant omitting their
explicit dependence on time in the notation. Also we note that for the power-based compensation of the harmonic part in $y(t)$, the estimation of the instantaneous triple $[\hat{Y}_0,\hat{A},\hat{\omega}](t)$ is assumed, while the phase parameter is not required.   

The discrete power-based control, introduced in
\cite{ruderman2024}, is given (for the second-order systems) by
\begin{equation}
u^{\prime}(t) = k \, \hat{\omega}^2 \hat{A}(t^*)
\label{eq:2:1}
\end{equation}
with
\begin{equation}
\hat{A}(t^*) \equiv \hat{A} \, \mathrm{sign} \bigl( y(t^*) -
\hat{Y}_0 \bigr), \label{eq:2:2}
\end{equation}
where $t^*$ denotes the time instant of the last extrema, i.e.
either minimum or maximum of the oscillating output subtracting the 
bias, i.e. $\bigl(y(t) - \hat{Y}_0 \bigr)$. Recall that the estimate
(correspondingly update) of the oscillations parameters $[\hat{Y}_0,\hat{A},\hat{\omega}]$ appears twice per period, cf. \cite{ruderman2024}. As shown in \cite{ruderman2024}, an
optimal gain (over one period of harmonic oscillations) is
\begin{equation}
k = \frac{\sqrt{3}}{2\pi}, \label{eq:2:3}
\end{equation}
that compensates for an undamped oscillation in $y(t)$. Moreover, for taking into account the phase lag of the feed-forward sub-dynamics in $G(s)$, cf. Fig. \ref{fig:controlstruct}, it was shown that the corresponding time delay factor 
\begin{equation}
T = \Bigl(2\pi + \arg \bigl[ \tilde{G}(j 2 \hat{\omega})\bigr] \Bigr) \,
\hat{\omega}^{-1} \label{eq:2:4}
\end{equation}
is required. Recall that $T$ shifts the control value $u^{\prime}$  by the negative phase lag with respect to the full period $2 \pi \omega^{-1}$ and, thus, synchronizes the control value $\bar{u}$ with an internal input to the double integrator inside of $G(s)$. Note that some uncertainties due to disturbances and (eventually) feedback propagation at the input of the double-integrator are not taken into account. Thus, only an approximative $\tilde{G} \approx G(s) s^2$ can be assumed. The overall gain-shaped and time-delay-synchronized control value is then given by
\begin{equation}
\bar{u}(t) = K \, u^{\prime}(t-T),
\label{eq:2:5}
\end{equation}
cf. Fig. \ref{fig:controlstruct}, where the gain factor $K$ must be inverse to the input-output power norm of the gain-reducing sub-dynamics $\tilde{y}(j\omega) = \tilde{G}(j\omega) u (j\omega)$. Since the power-based control $u^{\prime}$ is the rectangular shaped piecewise constant value $U$, the corresponding amplitude reduction is $\bigl| \tilde{G}(s) \bigr| U$. This requires the amplification factor to be  
\begin{equation}
1 < K < \bigl| \tilde{G}(j \omega) \bigr|^{-1},
\label{eq:2:6}
\end{equation} 
provided $\tilde{G}(s)$ has some low-pass transfer characteristics.

\begin{figure}[!h]
\centering
\includegraphics[width=0.95\columnwidth]{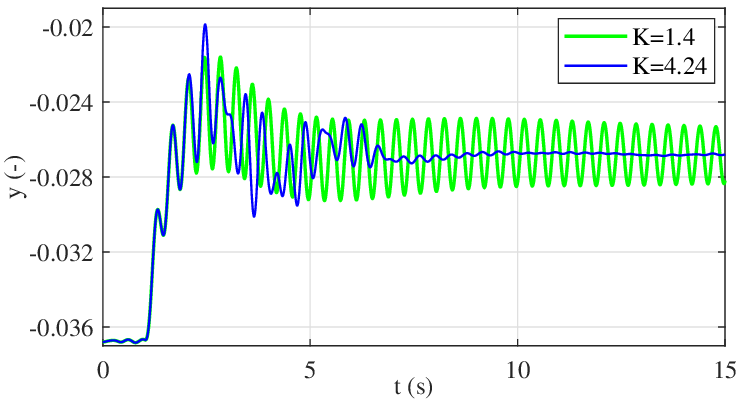}
\caption{Simulated response of the stabilized oscillating output for two boundary gain values $K$.} \label{fig:simresp}
\end{figure}

A numerically simulated response of the controlled system $G(s)$, with the simultaneous use of a destabilizing PI controller $\tilde{u}$ and the stabilizing power-based control $\bar{u}$ given by \eqref{eq:2:1},\eqref{eq:2:2},\eqref{eq:2:4},\eqref{eq:2:5} is exemplary shown in Fig. \ref{fig:simresp} for the sake of a better exposition. Note that the parameters of the system plant and feedback control are the same as the identified, correspondingly assigned, in the provided below experimental case study, cf. section \ref{sec:5}. Both boundary values of the gain amplification factor $K=\{1.4, \, 4.24\}$ are exemplified, cf. \eqref{eq:2:6}, while a lower boundary $> 1$ had to be assigned for stabilizing the otherwise unstable oscillations of $y(t)$, which is driven by the PI feedback $\tilde{u}$.   

Recall that for implementation of the above control strategy we need to identify online the amplitude of oscillations, the bias and the angular frequency. This is required to be sufficiently fast with regard to a half of the oscillations period (the sampling of the discrete control update).

\section{Estimation of biased harmonic}
\label{sec:3}

Following \cite{ahmed2022} and \cite{Ushirobira2023}, we can
observe that for (\ref{eq:output}) with $v\equiv0$ the following
relation is satisfied:
\begin{gather*}
y(t-3\tau)-y(t-2\tau)+y(t-\tau)-y(t)\\
=2\cos(\omega\tau)\left(y(t-2\tau)-y(t-\tau)\right)
\end{gather*}
for all $t\geq0$ and any time delay $\tau>0$. Therefore, the following
linear regression equation can be used for estimation of the
angular frequency $\omega$:
\begin{equation}
\tilde{y}(t)=\tilde{\varphi}(t)\theta_{0}+\tilde{v}(t),\label{eq:LR1}
\end{equation}
where
\begin{gather*}
\tilde{y}(t)=y(t-3\tau)-y(t-2\tau)+y(t-\tau)-y(t),\\
\tilde{\varphi}(t)=2\left(y(t-2\tau)-y(t-\tau)\right),\;\theta_{0}=\cos(\omega\tau),\\
\tilde{v}(t)=2\cos(\omega\tau)\left(v(t-\tau)-v(t-2\tau)\right)+v(t-3\tau)\\
-v(t-2\tau)+v(t-\tau)-v(t),
\end{gather*}
and the new noise $\tilde{v}$ is also bounded. Different
algorithms can be used to estimate the parameter $\theta_{0}$. In particular, the conventional gradient descent method can be applied:
\begin{equation}
\dot{\hat{\theta}}_{0}(t)=\gamma_{1}\tilde{\varphi}(t)\left(\tilde{y}(t)-\tilde{\varphi}(t)\hat{\theta}_{0}(t)\right)\label{eq:ls1}
\end{equation}
or its finite-time converging version \cite{Wang2020}:
\begin{equation}
\dot{\hat{\theta}}_{0}(t)=\gamma_{1}\tilde{\varphi}(t)\sqrt{|\tilde{y}(t)-\tilde{\varphi}(t)\hat{\theta}_{0}(t)|}\text{sign}\left(\tilde{y}(t)-\tilde{\varphi}(t)\hat{\theta}_{0}(t)\right).\label{eq:ls_ft}
\end{equation}
With an adaptation gain $\gamma_{1}>0$, then
\[
\hat{\omega}(t)=\tau^{-1}\arccos(\hat{\theta}_{0}(t))
\]
can be chosen as the estimate of the angular frequency.

Knowing the angular frequency, the problem of estimation of the bias
$Y_{0}$, the amplitude $A$ (and the phase $\phi$) can be also
formulated in the linear regression form:
\begin{align*}
y(t) & =\varphi^{\top}(t)\theta+v(t)\\
 & =\hat{\varphi}^{\top}(t)\theta+\hat{v}(t),
\end{align*}
for
\begin{gather*}
\varphi^{\top}(t)=[1\;\sin(\omega t)\;\cos(\omega t)],\\
\theta^{\top}=[Y_{0}\;A\cos(\phi)\;A\sin(\phi)],\\
\hat{\varphi}^{\top}(t)=[1\;\sin(\hat{\omega}(t)t)\;\cos(\hat{\omega}(t)t)],\\
\hat{v}(t)=v(t)+\left(\varphi(t)-\hat{\varphi}(t)\right)^{\top}\theta.
\end{gather*}
Then, a similar gradient descent algorithm can be utilized to estimate
$\theta$:
\begin{equation}
\dot{\hat{\theta}}(t)=\gamma_{2}\hat{\varphi}(t)(y(t)-\hat{\varphi}(t)\hat{\theta}(t))\label{eq:ls2}
\end{equation}
for $\gamma_{2}>0$, where
\begin{gather*}
\hat{Y}_{0}(t)=\hat{\theta}_{1}(t),\;\hat{A}(t)=\sqrt{\hat{\theta}_{2}^{2}(t)+\hat{\theta}_{3}^{2}(t)},\\
\hat{\phi}(t)=\arctan\left(\frac{\hat{\theta}_{3}(t)}{\hat{\theta}_{2}(t)}\right)
\end{gather*}
are the estimates of the remaining parameters in
(\ref{eq:output}).

In this case $\tau$, $\gamma_{1}$ and $\gamma_{2}$ are the
estimation algorithm parameters that have to be properly tuned.


\section{Convergence properties and parameters tuning}
\label{sec:4}

The gradient descent algorithms (\ref{eq:ls1}) or (\ref{eq:ls2})
guarantee convergence of the parameter estimates to their ideal
values in the noise-free setting if the regressors
($\tilde{\varphi}(t)$ or $\hat{\varphi}(t)$ in (\ref{eq:ls1}) or
(\ref{eq:ls2}), respectively) are persistently excited, cf. e.g.
\cite{Sastry1989,Efimov2015}. Recall that a bounded signal
$\hat{\varphi}(t)$ is called persistently excited
\cite{Morgan1977} if there exist $T^{\ast}>0$ and $\alpha>0$ such that
\[
\int_{t}^{t+T^{\ast}}\hat{\varphi}(s)\hat{\varphi}^{\top}(s)ds\geq\alpha
I
\]
for all $t\geq0$, where $I$ is the identity matrix of an appropriate
dimension. For the finite-time converging algorithm
(\ref{eq:ls_ft}), an interval excitation can be enough in such a
case \cite{Wang2020}, i.e., the above integral property is
verified for some $t\geq0$. In the presence of measurement noises,
all these algorithms ensure input-to-state stability of the
parameter identification errors with respect to the measurement
perturbations under persistence of excitation \cite{Wang2020}. The
latter property in the absence of noise can be easily verified for
$\tilde{\varphi}(t)$:
\begin{gather*}
\tilde{\varphi}(t)=2A\left(\sin(\omega t-2\omega\tau+\phi)-\sin(\omega t-\omega\tau+\phi)\right)\\
=-4A\sin(\frac{\omega}{2}\tau)\cos(\omega
t-\frac{3\omega}{2}\tau+\phi),
\end{gather*}
then for $T^{\ast}=\frac{2\pi}{\omega}$
\begin{gather*}
\int \limits_{t}^{t+T^{\ast}}\tilde{\varphi}^{2}(s)ds=16A^{2}\sin^{2}(\frac{\omega}{2}\tau)\int \limits_{t}^{t+T^{\ast}}\cos^{2}(\omega s-\frac{3\omega}{2}\tau+\phi)ds\\
=16A^{2}\frac{\pi}{\omega}\sin^{2}(\frac{\omega}{2}\tau),
\end{gather*}
and the excitation is maximal for
\[
\tau=\frac{\pi}{\omega},
\]
which provides indication for the choice of $\tau$. For
example, if $\underline{\omega}\leq\omega\leq\overline{\omega}$
for some known constants $\overline{\omega},\underline{\omega}$,
then
\[
\frac{\pi}{\overline{\omega}}\leq\tau\leq\frac{\pi}{\underline{\omega}}
\]
is a recommended interval. Similarly, performing straightforward
calculations, the persistence of the excitation of $\varphi(t)$ can be
checked. We may assume that if the signal-noise ratio in
(\ref{eq:output}) is reasonable, then the same conclusions keep
their meaning for the noisy scenario also. It is worth, however,
to highlight that the value of $\tau$ has to be also chosen to
provide a sufficient distinguishability between the values $y(t)$,
$y(t-\tau)$, $y(t-2\tau)$ and $y(t-3\tau)$ in comparison to the
noise, i.e., for values of $\tau$ sufficiently small, the useful
part of the signal, $\sin(\omega t+\phi)$, stays approximately the
same, while the noise may vary arbitrary in its limits, hence,
impacting strongly the identification.

For tuning the adaptation gains $\gamma_{1}$,
$\gamma_{2}$, we may recall the estimates obtained in
\cite{Efimov2015}, which indicate that depending on the excitation
level, there exist their optimal values providing the best
convergence rates and sensitivity to the noise. However, the excitation level may be difficult to determine before the experiment.

The performance of the convergence can be improved using the Dynamic
Regressor Extension and Mixing method (\cite{Aranovskiy2017}), which
allows the monotone convergence of identification errors with
regulated rate of convergence at the price of introduction of
additional filters. This method can also be used for finite-time estimation of the parameters vector
$\theta$, improving the quality of identification with respect
to the algorithm (\ref{eq:ls2}) but demanding, however, an additional computational
effort.

\section{Experimental case study}
\label{sec:5}

\subsection{Mechanical oscillatory setup} \label{sec:5:sub:1}

\begin{figure}[!h]
\centering
\includegraphics[width=0.45\columnwidth]{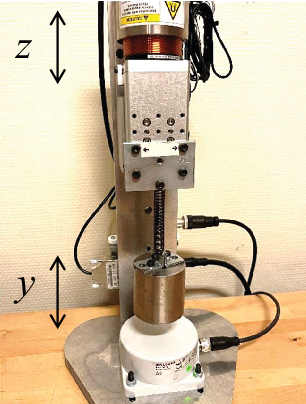} \hspace{4mm}
\includegraphics[width=0.38\columnwidth]{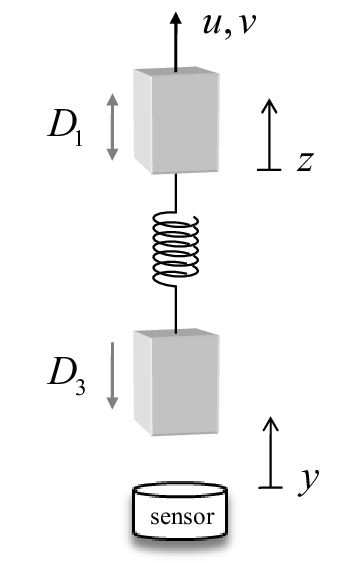}
\caption{Experimental oscillatory setup: laboratory view (left)
and equivalent mechanical scheme (right).} \label{fig:expsetup}
\end{figure}

The experimental control evaluation is accomplished on the
actuated mechanical setup (\cite{ruderman2022}) with the free
hanging oscillatory load, see Fig. \ref{fig:expsetup}.
The vertical motion with a constrained degree of freedom $z \in
[0,\, 0.021]$ m is induced by the voice-coil-based linear actuator
which receives the control values $u \in [0,\, 10]$ V. Both moving
masses are connected via an elastic spring which has the
relatively high stiffness but, at the same time, is subject to
hardening effects and uncertainties. The load relative
displacement $y(t)$ is the output value of interest and it is
measured contact-less by an inductive sensor located below.
Despite an axial alignment of both degrees of freedom $z$ and $y$
in series, cf. Fig. \ref{fig:expsetup} (right), the latter is
subject to an additional process noise. Recall that the free
hanging load is without any bearing and, thus, can undergo
additional degrees of freedom, even though
marginal. A real-time board provides the control and load
displacement signals with the sampling rate of $f_s = 2$ kHz. More
details on the system dynamics and parameters can be
found in e.g. \cite{ruderman2023}.

The known system model structure and nominal (also partially
identified) parameters are given by
\begin{equation}
\nu(s) = H(s)\, u(s) = \frac{3.2811}{0.0012 s + 1}\, u(s),
\label{eq:5:1}
\end{equation}
\begin{eqnarray}
\label{eq:5:2}
  \dot{x}(t) &=& A \, x(t) + B \,\nu(t) + D, \\
\nonumber y(t) &=& C \, x(t),
\end{eqnarray}
with the state vector $x = (\dot{z},z,\dot{y},y)^{\top}$ and
\begin{eqnarray}
\label{eq:5:3}
A &=& \left(%
\begin{array}{cccc}
-333.35 &  -333.33  &   0.015     & 333.33  \\
1         & 0     &     0         & 0         \\
0.012     & 266.66 &    -0.012    & -266.66 \\
0         & 0          &  1       & 0
\end{array}%
\right), \\
\nonumber B &=& (1.667, 0, 0, 0)^\top, \quad  C \; = \; (0, 0, 0, 1), \; \hbox { and} \\
\nonumber D &=& (-9.806 + 0.83 \, \mathrm{sign}(\dot{x}_1), 0,
-9.806, 0)^\top.
\end{eqnarray}
When compensating for the constant disturbance terms in $D$ (see
Fig. \ref{fig:expsetup} on the right), which are mainly due to the
gravity force, and neglecting the remaining Coulomb friction disturbance of
the actuator, the overall input-to-output transfer function of the
system is given by
\begin{equation}
G(s) = y(s) u(s)^{-1} = C (s I - A )^{-1} B \, H(s),
\label{eq:5:4}
\end{equation}
where $I$ is the $4 \times 4$ identity matrix, cf. Fig. \ref{fig:controlstruct}.

\subsection{Unstable PI feedback control}
\label{sec:5:sub:2}

\begin{figure}[!h]
\centering
\includegraphics[width=0.95\columnwidth]{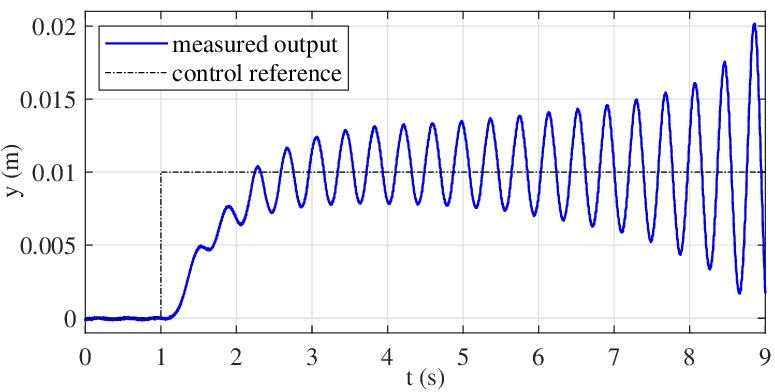}
\caption{Unstable load position response of PI-control.}
\label{fig:expres1}
\end{figure}

The output displacement $y(t)$ of the oscillatory passive load is
first controlled by a simple PI (proportional-integral) feedback
regulator
\begin{equation}
\tilde{u}(t) = 140 \bigl(R_1 - y(t)\bigr) + 170 \int \bigl(R_1 -
y(t)\bigr) dt + R_2. \label{eq:5:2:1}
\end{equation}
It is expected that the PI feedback control alone can neither
suppress the undesired output oscillations nor ensure the
stability of the closed control loop, considering that the loop
transfer function is of the fifth order. However, an integral control
action is still necessary for regulation towards the reference value
$R_1$, since multiple system uncertainties can not be captured by
the nominal plant model, like e.g. uncertainties of the mechanical
friction or force ripples of the voice-coil-motor. On the
contrary, the known total gravity force (induced by both the
actuator and the load masses) is feed-forward compensated by the constant
$R_2$, cf. \eqref{eq:5:2:1}. The measured unstable output response
of the load position is shown in Fig. \ref{fig:expres1}.

\begin{figure}[!h]
\centering
\includegraphics[width=0.95\columnwidth]{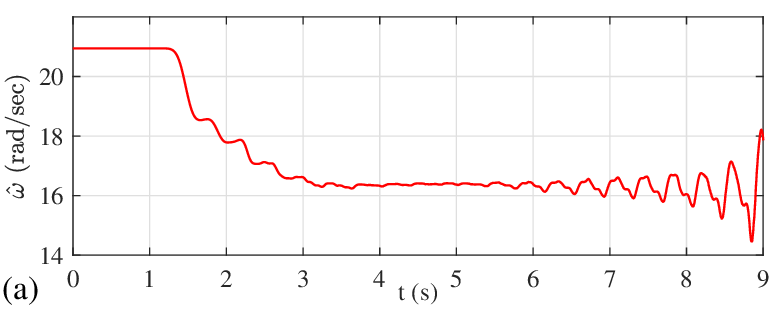}
\includegraphics[width=0.95\columnwidth]{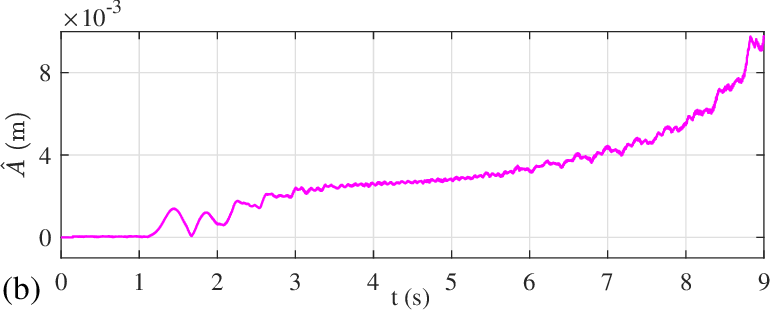}
\includegraphics[width=0.95\columnwidth]{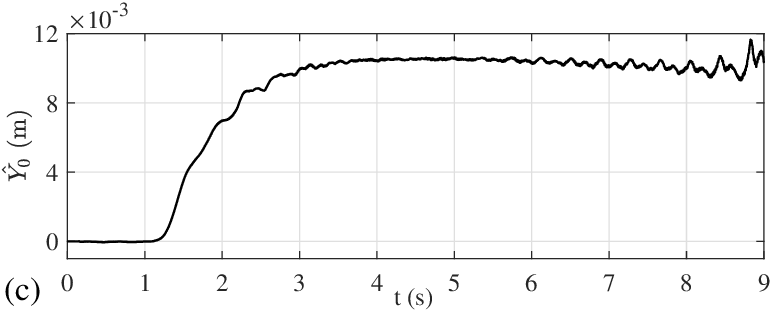}
\caption{Online estimation of harmonic parameters at unstable response:
frequency (a), bias (b), amplitude (c).} \label{fig:expestim}
\end{figure}

The parameters estimate is performed simultaneously on
the measured (diverging) oscillatory output signal. The tuned parameters of
the estimation algorithm (cf. sections \ref{sec:3}, \ref{sec:4})
are $\tau = 0.075$ sec, and $\gamma_1 = 1.5 \times 10^5$,
$\gamma_2 = 10^6$. The online parameters estimation over time is
shown for $\hat{\omega}$, $\hat{A}$, and $\hat{Y_0}$ in Fig.
\ref{fig:expestim} (a), (b), and (c), respectively.
One can recognize that once the $y(t)$ signal starts to diverge
faster (at time about $t > 8$ sec, cf. Fig. \ref{fig:expres1}),
the parameters estimate loses the convergence, in particular
visible for $\hat{\omega}(t)$. That case, the parameters estimate
becomes inappropriate for a power-based control and can, in the worst
case, even accelerate destabilization of the output. It seems logical that for
an effective use of the online parameters estimation in combination with the
power-based control, both the convergence of the estimate and the
suppression of oscillations should occur within a few periods.

\subsection{Unstable PI feedback control with power-based control}
\label{sec:5:sub:3}

\begin{figure}[!h]
\centering
\includegraphics[width=0.95\columnwidth]{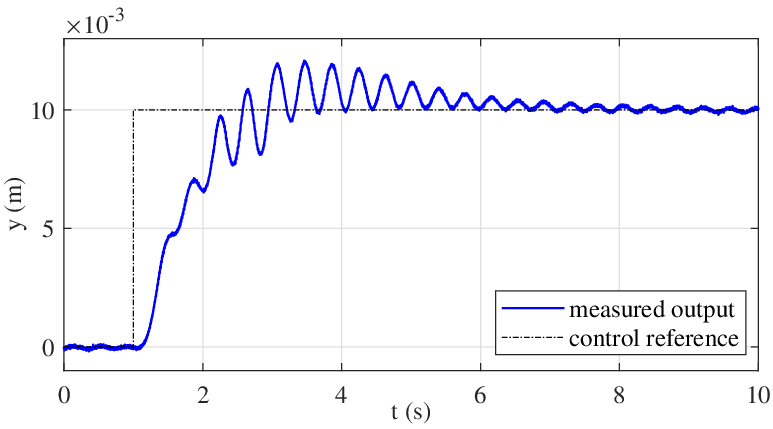}
\caption{Stabilized load position response of PI-control augmented
by power-based control at $t=2.5$ sec.} \label{fig:expres2}
\end{figure}

\begin{figure}[!h]
\centering
\includegraphics[width=0.95\columnwidth]{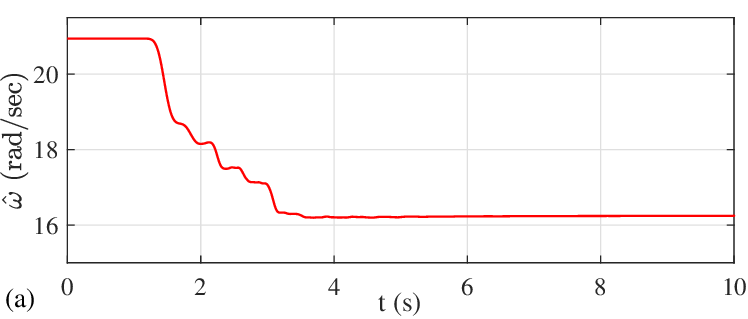}
\includegraphics[width=0.95\columnwidth]{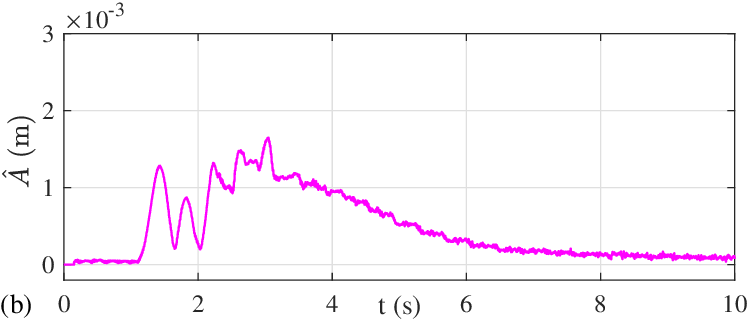}
\includegraphics[width=0.95\columnwidth]{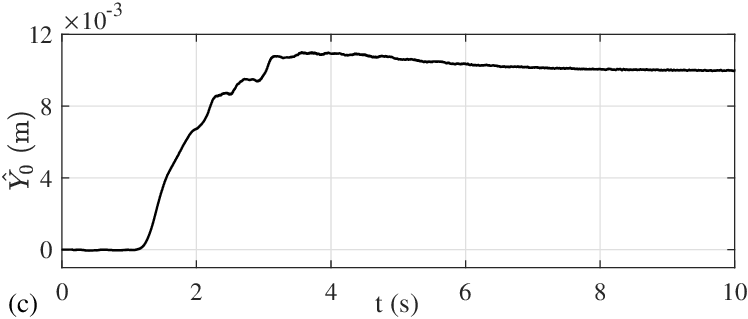}
\caption{Online estimation of harmonic parameters at stable response:
frequency (a), bias (b), amplitude (c).}
\label{fig:expestimcomp}
\end{figure}

Next, the power-based control \eqref{eq:2:5} with \eqref{eq:2:4}, \eqref{eq:2:1} and \eqref{eq:2:2} is switched on at time $t=2.5$ sec, thus augmenting the
otherwise unstable PI regulator \eqref{eq:5:2:1}, cf. Fig.
\ref{fig:controlstruct}. The set parameters of the estimation
algorithm are the same as above, cf. section \ref{sec:5:sub:2}. The determined feed-forward plant sub-dynamics (based on the identified model) is 
\begin{equation}
\tilde{G}(s) = \frac{0.1544 s + 3432}{0.002824 s^3 + 3.295 s^2 + 785.5 s + 784.5}, 
\label{eq:5:3:1}
\end{equation}
that results in the gain upper bound, cf. \eqref{eq:2:6},
$$
\bigl| \tilde{G}(j \omega) \bigr|^{-1} = 4.24.
$$
The properly tuned gain, assigned in the real-time control experiments, is $K = 2.4$.
The measured load position response, stabilized by the power-based
control is shown in Fig. \ref{fig:expres2}.
Also here, the online parameters estimation over the time are shown for
the sake of completeness in Fig. \ref{fig:expestimcomp}, for
$\hat{\omega}$, $\hat{A}$, and $\hat{Y_0}$ in (a), (b), and (c),
respectively.

\section{Conclusions}
\label{sec:6}

Pursuing the goal of reduction of the communication efforts within a control loop, particularly when compensating for marginally damped or slowly diverging output oscillations with non-collocated actuation and load sensing, a new method called discrete power-based control was recently introduced in \cite{ruderman2024}. The control commutates twice per oscillations period and for its effective application, the power-based compensation scheme necessitates the knowledge of several key parameters, as the instantaneous angular frequency, the amplitude, and the bias of the harmonic signal. These have to be reliable and rapidly estimated online. This paper extended the capabilities of the power-based control by incorporating a finite-time estimation of the required parameters, following the ideas of \cite{Wang2020,ahmed2022}. In addition, an improved analytic calculation of the impulse weighting factor was proposed, enhancing the control's effectiveness. The power-based oscillation control with online estimation of the harmonic parameters was experimentally evaluated. The tests were conducted on a fifth-order actuator system with a free-hanging load subjected to gravity and measurement noise. The evaluation provided insights into the applicability and performance of the control method in practical scenarios.

\section*{Acknowledgement}
This research was partially funded by AURORA mobility programme (RCN grant 340782). The first author acknowledges also the financial support by NEST (Network for Energy Sustainable Transition) foundation during the sabbatical year at Polytechnic University of Bari.

\bibliography{references}             

\end{document}